\documentclass[sigconf,nonacm]{acmart}

\usepackage{booktabs} 
\usepackage{makecell} 
\usepackage{bbm}
\usepackage{enumitem}
\usepackage{CJKutf8}
\usepackage[most]{tcolorbox}
\usepackage{tabularx}
\usepackage{multirow}
\usepackage{ragged2e}

\definecolor{promptblue}{RGB}{239,246,255}
\definecolor{promptborder}{RGB}{96,139,193}
\definecolor{promptlabel}{RGB}{40,85,140}

\newtcolorbox{promptbox}{
  colback=promptblue,
  colframe=promptborder,
  boxrule=0.6pt,
  arc=2pt,
  left=6pt,
  right=6pt,
  top=5pt,
  bottom=5pt,
  breakable
}

\AtBeginDocument{%
  }

\begin{document}

\title{MIPR: A Benchmark for Reasoning-Intensive Multimodal Intent-to-Product Retrieval with Multi-Category Solutions}

\author{Qiang Li}
\authornote{Qiang Li, Xuan Lu, Jian Ruan, and Manxi Lin contributed equally to this work.}
\affiliation{%
  \institution{Tongji University}
  \city{Shanghai}
  \country{China}
}

\author{Xuan Lu}
\authornotemark[1]
\affiliation{%
  \institution{Taobao \& Tmall Group of Alibaba}
  \city{Hangzhou}
  \country{China}
}

\author{Jian Ruan}
\authornotemark[1]
\affiliation{%
  \institution{Tongji University}
  \city{Shanghai}
  \country{China}
}

\author{Manxi Lin}
\authornotemark[1]
\affiliation{%
  \institution{Taobao \& Tmall Group of Alibaba}
  \city{Hangzhou}
  \country{China}
}

\author{Jiacheng Chen}
\affiliation{%
  \institution{Taobao \& Tmall Group of Alibaba}
  \city{Hangzhou}
  \country{China}
}

\author{Wen Shen}
\affiliation{%
  \institution{Tongji University}
  \city{Shanghai}
  \country{China}
}

\author{Zhihua Wei}
\affiliation{%
  \institution{Tongji University}
  \city{Shanghai}
  \country{China}
}

\author{Xiaoyu Shen}
\affiliation{%
  \institution{Eastern Institute of Technology, Ningbo}
  \city{Ningbo}
  \country{China}
}

\author{Dan Ou}
\affiliation{%
  \institution{Taobao \& Tmall Group of Alibaba}
  \city{Hangzhou}
  \country{China}
}

\author{Xiangheng Kong}
\affiliation{%
  \institution{Taobao \& Tmall Group of Alibaba}
  \city{Hangzhou}
  \country{China}
}

\author{Junjun Zheng}
\authornote{Corresponding author: fangcheng.zjj@alibaba-inc.com.}
\affiliation{%
  \institution{Taobao \& Tmall Group of Alibaba}
  \city{Hangzhou}
  \country{China}
}

\renewcommand{\shortauthors}{Li et al.}

\begin{abstract}
Real-world e-commerce queries often express user intent without naming a target product, and a single intent may be fulfilled by products from multiple categories. However, existing multimodal retrieval benchmarks typically assume a relatively well-defined target, leaving retrieval for intents with multi-category solutions less explored. To fill this gap, we introduce MIPR, a benchmark for Multimodal Intent-to-Product Retrieval comprising 3,522 multimodal queries and a shared corpus of 30,731 products. MIPR features reasoning-intensive retrieval, multi-category solutions, and fine-grained relevance labels. Experiments show that the best embedding-based retriever achieves 49.23\% nDCG@10 on the full query set, and 55.86\% solution-category recall@10 on queries with multiple solution categories. Reranking improves these scores to 60.11\% and 67.00\%, respectively, yet substantial gaps to the candidate-pool oracle remain. These results establish MIPR as a challenging and diagnostic benchmark for evaluating multimodal intent-to-product retrieval. The data and code are publicly available at https://huggingface.co/datasets/leequeue/MIPR.

\end{abstract}




\maketitle

\section{Introduction}
\label{sec:intro}

\begin{figure*}[t]
  \centering
  \includegraphics[width=\linewidth]{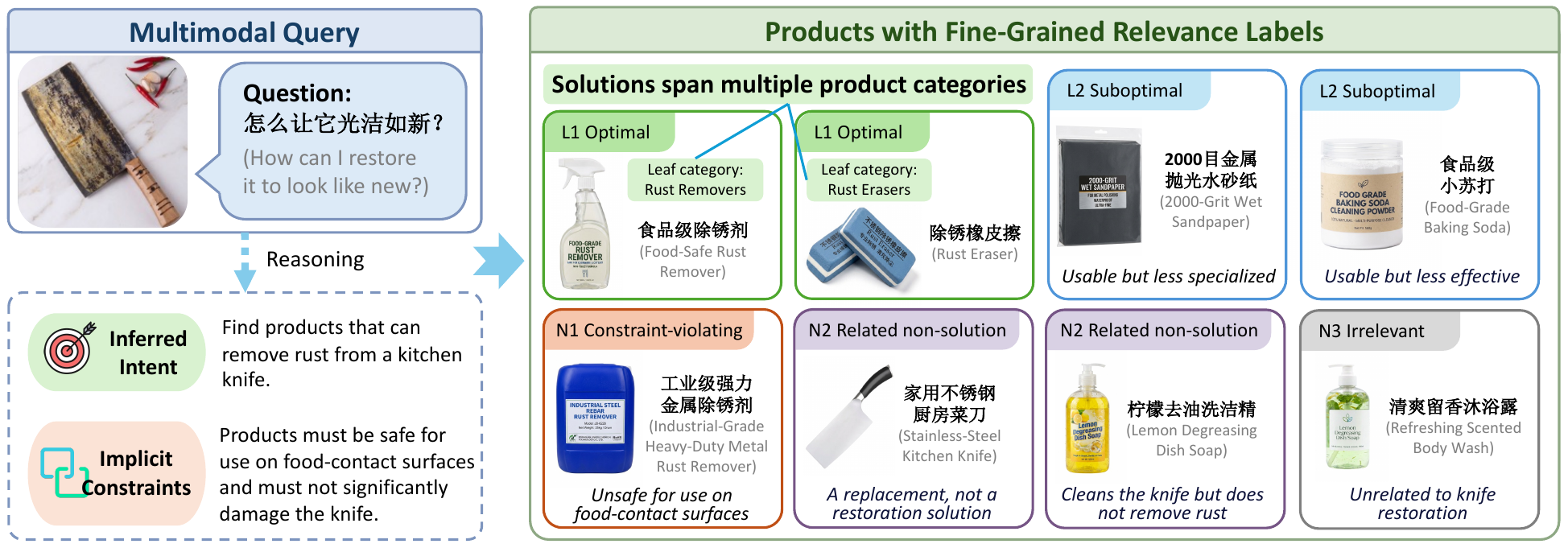}
  \caption{An example from MIPR. Each query combines an image and a question to express a user intent without specifying a target product, requiring models to understand the intent and reason about which products can fulfill it while satisfying any applicable constraints. Products are assigned fine-grained relevance labels, and the leaf-level categories of L1 products are defined as solution categories. Text is translated and lightly edited for clarity.}
  \Description{An example query about removing rust from a kitchen knife, together with products from multiple solution categories and their relevance labels.}
  \label{fig:intro}
\end{figure*}



Multimodal retrieval has expanded from text-to-image retrieval~\cite{clip,li2022blip} to composed image retrieval, which retrieves a target based on a reference image and a textual modification~\cite{wu2021fashioniq,liu2021cirr}, and more recently to reasoning-intensive retrieval, where relevance must be inferred through reasoning rather than direct semantic matching~\cite{li2026rmir,zhang2025mrmr}. Despite these advances, existing retrieval paradigms are typically target-centric: a query either directly specifies the target or points to a relatively well-defined target concept. Real-world e-commerce search, however, often begins with a need rather than a target. Users may upload an image of their situation and ask how to fulfill a need or solve a problem without naming any desired product. Retrieving useful products for such queries requires reasoning from the user's intent to possible solutions. Importantly, a single intent may admit solutions from multiple product categories that differ substantially in appearance and function.

Consider a shopper who uploads a photo of a rusty kitchen knife and asks, ``How can I restore it to look like new?'' Rather than pointing to a single target, this need admits solutions from multiple product categories. A food-safe rust remover and a knife rust eraser offer effective solutions from different categories, whereas fine-grit sandpaper is applicable but less specialized. Other candidates may appear relevant but are not valid solutions: an industrial rust remover may remove the rust but violate implicit food-safety constraints, while a new knife is related to the query but fails to fulfill the user's intent. Thus, evaluating such candidates requires fine-grained relevance judgments based on whether and how well they fulfill the underlying intent.

Existing benchmarks do not adequately support the evaluation of this retrieval setting. Semantic-matching benchmarks such as M-BEIR~\cite{wei2024uniir-mbeir}, MERIT~\cite{chow2026merit}, and MCMR~\cite{lu2026mcmr} primarily evaluate retrieval based on direct semantic correspondence between queries and relevant items. Reasoning-intensive benchmarks such as RMIR~\cite{li2026rmir} and MRMR~\cite{zhang2025mrmr} require reasoning toward an unstated target, but their relevant items typically belong to a single target category and use binary relevance labels. To our knowledge, no existing benchmark jointly supports reasoning-intensive retrieval, multi-category solutions, and fine-grained relevance labels.

To fill this gap, we introduce MIPR (Multimodal Intent-to-Product Retrieval), a benchmark grounded in real Chinese e-commerce search logs, comprising 3,522 multimodal queries and a shared corpus of 30,731 products. Each query is an image--question pair that expresses a user need without directly specifying a target product, requiring models to reason from the multimodal intent to suitable products that may span multiple categories. MIPR jointly captures three properties: (i) reasoning-intensive retrieval, where target products must be inferred rather than directly matched; (ii) multi-category solutions, where a query can be fulfilled by products from multiple categories; and (iii) fine-grained relevance labels, including L1 optimal, L2 suboptimal, N1 constraint-violating, N2 related non-solution, and N3 irrelevant. We construct MIPR through a documented pipeline combining multi-model labeling, human verification, and pooling-based judgment expansion, and release the pipeline to support future benchmark construction. Figure~\ref{fig:intro} illustrates a representative query, its solution categories, and candidate products with different relevance labels.

We evaluate a broad suite of state-of-the-art multimodal embedding retrievers and rerankers on MIPR. The results show that existing models struggle with both relevance ranking and multi-category solution coverage. The best embedding retriever achieves an nDCG@10 of only 49.23\%, places an L1 product at rank 1 for fewer than half of the queries, and achieves only 55.86\% solution-category recall@10 on queries with multi-category solutions. Reranking substantially improves performance, increasing nDCG@10 by 10.88 points and solution-category recall@10 to 67.00\%. Nevertheless, the best reranker remains 25.93 points below the candidate-pool oracle in nDCG@10 and 13.22 points below it in solution-category recall@10, highlighting substantial room for improvement. Together, these results establish MIPR as a challenging and diagnostic benchmark for multimodal intent-to-product retrieval.

\section{Related Work}
\label{sec:related}

\begin{table}[t]
  \centering
  \caption{Comparison of retrieval benchmarks. \emph{Modal.} denotes the query$\rightarrow$target modality (I: image; T: text; $+$: jointly used; $/$: task-dependent; Mixed: multiple target modalities). \emph{Reason.} indicates whether relevant targets must be inferred rather than directly specified by the query. \emph{Multi-Cat.} indicates whether the benchmark explicitly identifies and evaluates multiple solution categories for a single query. \emph{FG-Rel.} indicates whether graded or typed relevance labels beyond binary judgments are provided. Among the compared benchmarks, MIPR is the only one supporting all three properties.}
  \label{tab:benchmark_comparison}
  \setlength{\tabcolsep}{6pt}
  \small
  \begin{tabular}{lcccc}
    \toprule
    Benchmark & Modal. & Reason. & Multi-Cat. & FG-Rel. \\
    \midrule
    FashionIQ~\cite{wu2021fashioniq} & I+T$\rightarrow$I & \(\times\) & \(\times\) & \(\times\) \\
    CIRR~\cite{liu2021cirr} & I+T$\rightarrow$I & \(\times\) & \(\times\) & \(\times\) \\
    M-BEIR~\cite{wei2024uniir-mbeir} & I/T$\rightarrow$I/T & \(\times\) & \(\times\) & \(\times\) \\
    MERIT~\cite{chow2026merit} & I+T$\rightarrow$I+T & \(\times\) & \(\times\) & \(\times\) \\
    MCMR~\cite{lu2026mcmr} & I+T$\rightarrow$I+T & \(\times\) & \(\times\) & \(\times\) \\
    \midrule
    BRIGHT~\cite{su2025bright} & T$\rightarrow$T & \(\checkmark\) & \(\times\) & \(\times\) \\
    RMIR~\cite{li2026rmir} & I+T$\rightarrow$I & \(\checkmark\) & \(\times\) & \(\times\) \\
    MRMR~\cite{zhang2025mrmr} & I+T$\rightarrow$I+T & \(\checkmark\) & \(\times\) & \(\times\) \\
    MR\textsuperscript{2}-Bench~\cite{zhou2025mr2bench} & I+T$\rightarrow$Mixed & \(\checkmark\) & \(\times\) & \(\times\) \\
    \midrule
    \textbf{MIPR (ours)} & I+T$\rightarrow$I+T & \(\checkmark\) & \(\checkmark\) & \(\checkmark\) \\
    \bottomrule
  \end{tabular}
\end{table}

\textbf{Multimodal Retrievers.}
Early multimodal retrieval models such as CLIP~\cite{clip} and BLIP~\cite{li2022blip} learn shared image--text representations for cross-modal matching. Recent work adapts multimodal large language models into general-purpose embedding retrievers that process interleaved image--text inputs and follow retrieval instructions~\cite{jiang2024vlm2vec,zhang2025gme,li2026qwen3-vl-embedding,lin2025mmembed,meng2025vlm2vecv2}. Other methods improve fine-grained matching under multiple query conditions~\cite{chow2026merit} or incorporate reasoning into embedding retrieval~\cite{lan2025ume-r1}. However, these methods have not been systematically evaluated for reasoning-intensive retrieval with multi-category solutions and fine-grained relevance labels. We evaluate representative embedding models on MIPR in Section~\ref{sec:experiments}.

\textbf{Multimodal Retrieval Benchmarks.}
Table~\ref{tab:benchmark_comparison} compares representative semantic-matching and reasoning-intensive benchmarks with MIPR along three dimensions: reasoning-intensive retrieval, multi-category solutions, and fine-grained relevance labels. \emph{Semantic-matching benchmarks} evaluate retrieval based primarily on content or attributes specified in the query. M-BEIR~\cite{wei2024uniir-mbeir} unifies diverse cross-modal retrieval tasks, FashionIQ~\cite{wu2021fashioniq} and CIRR~\cite{liu2021cirr} use a reference image and a textual modification to characterize the desired target, and MERIT~\cite{chow2026merit} and MCMR~\cite{lu2026mcmr} evaluate fine-grained matching under multiple query conditions. These tasks require detailed multimodal understanding and composition, but relevance is primarily determined by matching query-specified content or attributes. \emph{Reasoning-intensive benchmarks} require models to infer relevant targets that are not directly specified by the query. BRIGHT~\cite{su2025bright} studies this setting in text retrieval, while RMIR~\cite{li2026rmir}, MRMR~\cite{zhang2025mrmr}, and MR\textsuperscript{2}-Bench~\cite{zhou2025mr2bench} extend it to multimodal queries. However, their relevant items typically belong to a single target category and are annotated with binary relevance labels. They therefore do not capture user needs that admit solutions from multiple product categories or distinguish how well a product fulfills the underlying intent. In contrast, MIPR jointly supports reasoning-intensive retrieval, multi-category solutions, and fine-grained relevance labels.

\section{The MIPR Benchmark}
\label{sec:benchmark}

MIPR is constructed from real Chinese e-commerce search logs to evaluate retrieval from multimodal user intents to products across multiple solution categories. Each query consists of an image providing situational context and a question expressing the user's need or concern. Products are drawn from a shared corpus and annotated with fine-grained relevance labels. We first formulate the task (Section~\ref{sec:task_formulation}) and then describe benchmark construction and statistics in Sections~\ref{sec:pipeline} and~\ref{sec:statistics}, respectively.

\begin{figure*}[t]
  \centering
  \includegraphics[width=\linewidth]{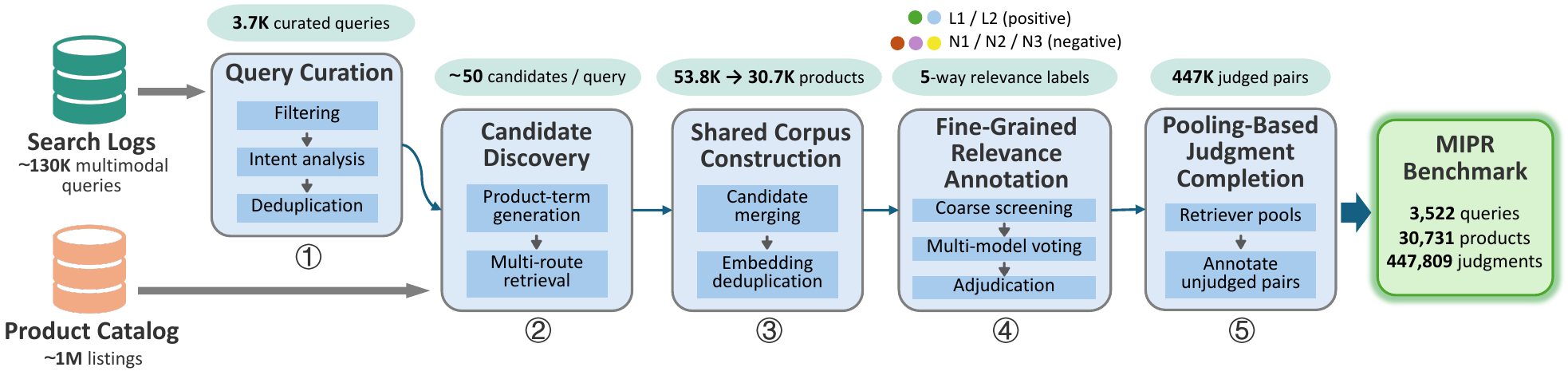}
  \caption{Overview of the MIPR construction pipeline. Starting from real multimodal search queries and a large product catalog, we curate eligible queries, discover candidate products through complementary retrieval routes, construct a deduplicated shared corpus, assign five-way relevance labels, and complete the judgments through retrieval-based pooling.}
  \Description{Overview of the five-stage MIPR construction pipeline. Starting from real multimodal search queries and a large product catalog, we curate eligible queries, discover candidate products through complementary retrieval routes, construct a deduplicated shared corpus, assign five-way relevance labels, and complete the judgments through retrieval-based pooling.}
  \label{fig:pipeline}
\end{figure*}

\subsection{Task Formulation}
\label{sec:task_formulation}

Given a multimodal query $q=(q_I,q_T)$, where $q_I$ provides situational context and $q_T$ expresses the user's need or concern, the task is to rank products from a shared corpus $\mathcal{P}=\{p_1,\ldots,p_N\}$ according to how well they fulfill the intent jointly implied by the two modalities. Each product consists of an image and a textual description. Unlike target-centric retrieval, a query in MIPR may be fulfilled by products from multiple categories.

\textbf{Five-way relevance labels.}
We distinguish two levels of positive relevance and three types of negative relevance:
\begin{itemize}[nosep,leftmargin=*]
    \item \textbf{L1 (Optimal):} products that directly and effectively fulfill the user's intent while satisfying all applicable constraints;
    \item \textbf{L2 (Suboptimal):} products that provide a usable but less effective, less specialized, or auxiliary solution;
    \item \textbf{N1 (Constraint-violating):} products that could otherwise fulfill the intent but violate a critical explicit or implicit safety, compatibility, or applicability constraint;
    \item \textbf{N2 (Related non-solution):} products that are visually, semantically, or categorically related to the query context but do not fulfill the intent; and
    \item \textbf{N3 (Irrelevant):} products that neither fulfill the intent nor constitute an N1 or N2 negative.
\end{itemize}

\textbf{Solution categories.}
Because free-form solution boundaries are difficult to define and annotate consistently, we use Taobao's existing product taxonomy as an operational proxy. Each product is associated with a three- or four-level hierarchical category path, whose leaf category serves as the evaluation unit. This platform-defined taxonomy provides consistent and measurable units while preserving meaningful functional distinctions among products.

For each query $q$, we define its annotated solution-category set as
\begin{equation}
\mathcal{S}(q)=\{\operatorname{leaf}(p)\mid y(q,p)=\mathrm{L1}\},
\end{equation}
where $\operatorname{leaf}(p)$ denotes the leaf category of product $p$, and $y(q,p)$ denotes its relevance label. The number of annotated solution categories is $|\mathcal{S}(q)|$. Because relevance judgments are obtained through benchmark construction and pooling, $\mathcal{S}(q)$ represents the identified solution categories rather than an exhaustive set of all possible real-world solutions. We use this set to compute solution-category recall in Section~\ref{sec:experimental_setup}.

\subsection{Benchmark Construction Pipeline}
\label{sec:pipeline}


MIPR is constructed from two independently collected Taobao data sources: approximately 130K multimodal queries sampled from real search logs and a catalog of approximately 1M product listings. As shown in Figure~\ref{fig:pipeline}, construction proceeds in five stages. We describe each stage below, together with its intermediate scale and human validation where applicable. Additional implementation details, validation results, and prompts used in model-assisted stages are provided in Appendices~\ref{app:pipeline_validation} and~\ref{app:prompts}.

\subsubsection{Query Curation}
\label{sec:query_curation}

MIPR targets queries that express a product-fulfillable intent through the combination of an image and a question, without directly specifying the target product. Real search logs, however, are of mixed quality and contain many queries that do not meet this requirement. We therefore curate queries that require joint multimodal understanding, provide sufficient information for product selection, and cannot be resolved through direct matching to an explicitly specified target. Details are provided in Appendix~\ref{app:query_validation}.

\textbf{Filtering and intent analysis.}
We first apply conservative rules to remove clearly ineligible queries, followed by a lightweight model (Qwen3.5-9B) that filters additional obvious mismatches, reducing the initial pool to approximately 30K queries. To derive a consistent representation of each remaining query, Claude Sonnet 4.6 produces a structured query analysis comprising an image description, the inferred user intent, and explicit product-selection constraints. This analysis is reused throughout the subsequent construction stages.

GLM-5.2 and DeepSeek-V4-Pro~\cite{xu2026deepseekv4} then independently assess query eligibility based on both the original image--question pair and the structured analysis. Only queries accepted by both models are retained, resulting in approximately 9.8K queries. Human validation shows that 96.5\% of 200 sampled structured analyses correctly capture both the visual content and the user intent, while only 1.5\% of the corresponding queries are rated as unsuitable or borderline, supporting the reliability of both the structured analyses and the query-selection procedure.

\textbf{Redundancy reduction.}
Search logs contain many queries that differ in surface form while expressing nearly identical product needs. To reduce such redundancy, we embed the inferred intents, cluster them using a cosine-similarity threshold of 0.7, and retain one representative from each cluster. This step yields approximately 3.7K queries for the subsequent construction stages.

\subsubsection{Candidate Product Discovery}
\label{sec:product_term_gen}

For each curated query, we discover candidate products in two steps: first generating product terms that cover plausible solution types, and then retrieving concrete products through multiple complementary routes.

\textbf{Product-term generation.}
To reach solutions spanning multiple categories, we adopt a two-stage enumeration-and-reflection procedure using Qwen3.6-27B. The model receives the original question and the structured query analysis, including the visual context, inferred intent, and product-selection constraints. It first enumerates product terms from four perspectives: \emph{specialized} products designed for the need, \emph{substitutes} that achieve a similar outcome, \emph{complementary} products that support a primary solution, and \emph{prerequisites} required for a solution, where the latter two mainly contribute auxiliary solutions. The model then reflects on the initial enumeration to recover omitted solution types and remove redundant or off-topic terms, retaining at most six product terms per query. Human validation on 50 queries shows that the reflection step increases coverage of human-identified valid product categories from 81\% to 89\%. Details are provided in Appendix~\ref{app:candidate_validation}.

\textbf{Multi-route retrieval.}
To avoid restricting candidate discovery to the generated product terms, we combine term-based retrieval with direct intent-based retrieval. For each generated term, we retrieve candidate products using BM25, a fused ranking produced by two text embedding models, and a vision--language embedding model. In parallel, we directly use the inferred query intent to retrieve products through the text and vision--language embedding routes. Candidates from all routes are merged by product ID and deduplicated, resulting in approximately 50 candidate products per query on average. The retriever configurations, fusion procedure, and retrieval depths are detailed in Appendix~\ref{app:candidate_validation}.

\subsubsection{Shared Corpus Construction}
\label{sec:corpus_construction}

Rather than evaluating against the full catalog of approximately 1M products, we construct a shared corpus from the union of the candidates discovered for all curated queries. This design concentrates annotation effort on a broad but tractable set of potentially relevant products, while the shared corpus preserves cross-query competition: products introduced through one query remain retrievable for every other query. Merging all query-specific candidate sets produces an initial corpus of 53,786 unique products.

We further reduce near-duplicate product variants, which would otherwise dominate retrieval results without meaningfully increasing solution diversity. Specifically, we encode each product's image and title using Qwen3-VL-Embedding-8B~\cite{li2026qwen3-vl-embedding} and cluster products within each leaf category. Each cluster is represented by the product closest to its centroid, and query--candidate pairs involving other cluster members are redirected to this representative, yielding a shared corpus of 30,731 products. Finally, Claude Sonnet 4.6 generates an image description for each retained product for use in subsequent relevance annotation. Clustering thresholds and detailed statistics are provided in Appendix~\ref{app:corpus_validation}.

\subsubsection{Fine-Grained Relevance Annotation}
\label{sec:relevance_annotation}

We assign relevance labels to the discovered query--candidate pairs through a two-stage procedure, including coarse screening and fine-grained annotation.

\textbf{Coarse screening.}
As each query is associated with a large number of candidates, we first use a locally deployed Qwen3.6-27B model to perform binary relevance screening. This stage is designed to remove clearly irrelevant products while retaining borderline cases for fine-grained annotation. Rejected candidates are assigned N3, as coarse screening does not distinguish among negative labels. Human inspection of all candidates rejected for 50 randomly sampled queries finds that only 4.4\% should have received a non-N3 label. Most screening errors fall at the N2--N3 boundary.

\textbf{Fine-grained annotation.}
The remaining candidates are independently labeled by Qwen3.7-Max, GLM-5.2, and DeepSeek-V4-Pro~\cite{xu2026deepseekv4} according to the five-way relevance schema defined in Section~\ref{sec:task_formulation}. Each model receives the query's image description, original question, and inferred intent, together with the candidate product's image description, title, and category path. We aggregate the three labels by majority vote, and Claude Sonnet 4.6 adjudicates cases in which the disagreement crosses the positive--negative boundary or all three models assign different labels.

For validation, Annotator A labels all candidates associated with 50 randomly sampled queries, while Annotator B independently labels those associated with ten of these queries. On this double-annotated subset, they achieve 87.8\% exact agreement under the five-way schema before reconciliation, with most disagreements occurring at the N2--N3 boundary. Against the resulting human reference, the annotation pipeline achieves 81.3\% exact agreement and 81.4\% macro-F1 under the five-way schema. For L1, it achieves 85.8\% precision and 83.5\% recall; for N1, it achieves 88.2\% precision and 85.4\% recall. Full per-label results, annotation guidelines, and model-disagreement statistics are reported in Appendix~\ref{app:screening_validation}.

\subsubsection{Pooling-Based Judgment Expansion}
\label{sec:pooling}

Because products discovered for all queries are merged into a shared corpus, a product relevant to a query may have entered the corpus through another query's candidate set and thus remain unjudged for that query. Following standard pooling practice in IR~\cite{voorhees2005trec}, we construct additional judgment pools using the ten multimodal embedding models evaluated in Section~\ref{sec:experimental_setup}, with each model contributing its top-20 products per query. Every previously unjudged query--product pair in the resulting pools is annotated using the procedure described in Section~\ref{sec:relevance_annotation}. These additional judgments are then merged with the initial annotations to form the released pooled qrels.

Pooling contributes 299,479 of the 447,809 released judgments, increasing the average number of judged L1 products per query from 3.54 to 4.99 and the average number of judged L1 solution categories from 2.72 to 3.16. Thus, approximately 29\% of the L1 judgments and 14\% of the judged L1 solution categories in the released qrels are identified through pooling. Following standard pooled-evaluation practice, query--product pairs outside the judgment pools are assigned zero gain during metric computation. Because the pooling systems overlap with the evaluated retrievers, we examine the sensitivity of model comparisons to pooling-based judgment expansion in Section~\ref{sec:analysis_pooling}. Retriever configurations and detailed pool statistics are provided in Appendix~\ref{app:pooling}.

\subsection{Dataset Statistics}
\label{sec:statistics}

\begin{table}[t]
\small
\centering
\caption{Distribution of fine-grained relevance labels in MIPR.}
\label{tab:label_dist}
\setlength{\tabcolsep}{10pt}
\begin{tabular}{lrr}
\toprule
Label & Count & Ratio \\
\midrule
L1 (optimal)              & 17,559  & 3.92\%  \\
L2 (suboptimal)           & 26,732  & 5.97\%  \\
N1 (constraint-violating) & 9,832   & 2.20\%  \\
N2 (related non-solution) & 93,197  & 20.81\% \\
N3 (irrelevant)           & 300,489 & 67.10\% \\
\midrule
Total                     & 447,809 & 100\%   \\
\bottomrule
\end{tabular}
\end{table}

After pooling and relevance annotation, we remove queries with no judged L1 products, for which the L1-centered metrics are undefined. We also remove queries with at least 20 judged L1 products, as manual inspection found these queries to be predominantly broad or underspecified, resulting in weakly discriminative relevance judgments. The final benchmark contains 3,522 queries over a shared corpus of 30,731 products, with no predefined training split. The pooled qrels comprise 447,809 judged query--product pairs, averaging 127.1 judgments per query.

Each query has an average of 4.99 judged L1 products spanning 3.16 solution categories, with a median of 3 and a maximum of 16 categories. Overall, 75.6\% of queries have judged L1 products from at least two solution categories, and 51.8\% have products from at least three. Across the benchmark, the L1 judgments cover 2,782 distinct leaf categories. Table~\ref{tab:label_dist} summarizes the distribution of the five relevance labels. Representative examples illustrating the label schema and multi-category solutions are provided in Appendix~\ref{app:dataset_examples}.

All data are collected and used in accordance with the platform's user agreements and privacy policies. Ethics, privacy, and data-compliance details are provided in Appendix~\ref{app:data_compliance}.

\section{Experiments}
\label{sec:experiments}

\begin{table*}[t]
\centering
\caption{Embedding retrieval results on MIPR. We report graded nDCG with gains of 2 for L1, 1 for L2, and 0 for all other products. We also report macro-averaged product-level recall over judged L1 products ($\mathrm{R}_{L1}$), L1 hit rate ($\mathrm{Hit}_{L1}$), and macro-averaged L1 solution-category recall ($\mathrm{SC\text{-}R}_{L1}$). $\mathrm{SC\text{-}R}_{L1}$ is evaluated on both the full query set $\mathcal{Q}$ and the subset of queries with multi-category solutions $\mathcal{Q}_{\mathrm{multi}}$. All metrics are reported on a 0--100 scale. Higher values are better. Best results are shown in \textbf{bold}, and second-best results are \underline{underlined}.}
\label{tab:retrieval_results}
\small
\setlength{\tabcolsep}{7pt}
\begin{tabular}{lccccccccccc}
\toprule
& & \multicolumn{3}{c}{nDCG}
& \multicolumn{3}{c}{$\mathrm{R}_{L1}$}
& \multicolumn{2}{c}{$\mathrm{Hit}_{L1}$}
& \multicolumn{2}{c}{$\mathrm{SC\text{-}R}_{L1}$} \\
\cmidrule(lr){3-5}
\cmidrule(lr){6-8}
\cmidrule(lr){9-10}
\cmidrule(lr){11-12}
Model
& Size
& @10 & @20 & @50
& @10 & @20 & @50
& @1 & @10
& @10 ($\mathcal{Q}$) & @10 ($\mathcal{Q}_{\mathrm{multi}}$) \\
\midrule

CORAL
& 3B
& 10.53 & 11.77 & 14.41
& 12.94 & 17.59 & 26.48
& 9.11 & 30.21
& 16.32 & 13.13 \\

GME-Qwen2VL
& 7B
& 39.48 & 42.54 & 47.70
& \underline{46.78} & \underline{58.37} & \underline{72.32}
& \underline{37.68} & \underline{78.39}
& \underline{54.12} & \underline{47.98} \\

LamRA-Ret
& 7B
& 26.91 & 29.62 & 34.18
& 32.94 & 42.91 & 56.12
& 24.47 & 62.01
& 39.26 & 33.36 \\

LLaVE
& 7B
& 34.52 & 38.14 & 43.62
& 36.30 & 48.56 & 64.52
& 27.91 & 71.32
& 43.98 & 39.87 \\

MM-Embed
& 8B
& 27.84 & 30.28 & 34.94
& 29.06 & 38.79 & 52.91
& 24.56 & 62.32
& 36.53 & 33.13 \\

Qwen3-VL-Embedding
& 8B
& \textbf{49.23} & \textbf{52.76} & \textbf{57.52}
& \textbf{53.98} & \textbf{66.92} & \textbf{79.61}
& \textbf{47.02} & \textbf{85.63}
& \textbf{61.33} & \textbf{55.86} \\

UME-R1 (Disc)
& 7B
& 38.71 & 42.19 & 47.70
& 42.37 & 54.90 & 69.58
& 34.78 & 76.77
& 49.88 & 44.36 \\

UME-R1 (Gen)
& 7B
& \underline{39.89} & \underline{43.24} & \underline{48.57}
& 43.38 & 55.76 & 69.87
& 36.26 & 77.29
& 51.00 & 45.54 \\

VLM2Vec-Qwen2VL
& 7B
& 13.80 & 15.81 & 19.22
& 17.49 & 24.81 & 35.65
& 11.75 & 37.51
& 21.38 & 17.38 \\

VLM2Vec-V2
& 2B
& 7.51 & 8.77 & 10.85
& 10.23 & 14.64 & 21.74
& 6.50 & 22.12
& 12.62 & 9.10 \\

\bottomrule
\end{tabular}
\end{table*}

\subsection{Experimental Setup}
\label{sec:experimental_setup}

We evaluate publicly available checkpoints on MIPR in a zero-shot setting. Models receive only the original query image and question, while each product is represented by its original image and title. Construction-time analyses and descriptions are not exposed. All retrievers operate on the same corpus.

\textbf{Embedding Retrievers.}
We evaluate ten embedding retrievers: CORAL~\cite{chow2026merit}, GME-Qwen2VL-7B~\cite{zhang2025gme}, LamRA-Ret~\cite{liu2025lamra}, LLaVE-7B~\cite{lan2025llave}, MM-Embed~\cite{lin2025mmembed}, UME-R1-7B~\cite{lan2025ume-r1}, Qwen3-VL-Embedding-8B~\cite{li2026qwen3-vl-embedding}, VLM2Vec-Qwen2VL-7B~\cite{jiang2024vlm2vec}, and VLM2Vec-V2~\cite{meng2025vlm2vecv2}. For UME-R1, we evaluate both its discriminative and generative variants, with the latter performing chain-of-thought reasoning before producing the query embedding. Each model independently encodes the multimodal queries and products, and ranks products by embedding similarity.

\textbf{Rerankers.}
We rerank the top-50 candidates returned by the strongest first-stage retriever, selected according to nDCG@10. We evaluate two general-purpose MLLMs, Qwen2-VL-7B-Instruct~\cite{wang2024qwen2vl} and Qwen3-VL-8B-Instruct~\cite{bai2025qwen3vl}, and two specialized rerankers, Qwen3-VL-Reranker-8B~\cite{li2026qwen3-vl-embedding} and Lychee-Rerank-MM~\cite{dai2025lychee}. Given the original multimodal query together with a candidate product's image and title, each model assesses the candidate's relevance to the user intent. We score each candidate using the probabilities obtained by normalizing the next-token logits for \texttt{yes} and \texttt{no}, without generating a full response, and rank candidates by the resulting \texttt{yes} probability. Implementation details are provided in Appendix~\ref{app:reranker_details}.

\textbf{Evaluation Metrics.}
Our primary metrics are nDCG, L1 product recall ($\mathrm{R}_{L1}$), and L1 hit rate (Hit$_{L1}$@$K$). We compute nDCG@$K$ using gains of 2, 1, and 0 for L1, L2, and all other products, respectively. This metric rewards valid solutions while prioritizing optimal products over suboptimal ones. $\mathrm{R}_{L1}$@$K$ measures the fraction of judged L1 products retrieved in the top $K$, averaged across queries, and thus evaluates optimal-product coverage. Hit$_{L1}$@$K$ measures the proportion of queries for which the top $K$ contains at least one L1 product.

To characterize category coverage, we report L1 Solution-Category Recall (SC-R$_{L1}$@$K$). For each query, this metric computes the fraction of judged L1 product categories represented by at least one retrieved L1 product in the top $K$. We report SC-R$_{L1}$@$K$ on both the full query set $\mathcal{Q}$ and the multi-category subset $\mathcal{Q}_{\mathrm{multi}}=\{q\in\mathcal{Q}\mid |\mathcal{S}(q)|\geq 2\}$, enabling a focused evaluation of category coverage for queries with at least two judged L1 solution categories. Together, Hit$_{L1}$ and SC-R$_{L1}$ distinguish retrieving at least one optimal product from covering optimal products across multiple categories. Additional diagnostic metrics measuring preference between optimal and suboptimal solutions, as well as the ranking of constraint-violating and related non-solution negatives, are introduced in Sections~\ref{sec:analysis_graded} and~\ref{sec:analysis_hardneg}.

\subsection{Embedding Retrieval Results}
\label{sec:embedding_results}

Table~\ref{tab:retrieval_results} presents the performance of embedding retrievers on MIPR. Overall, the task remains challenging for existing models. Qwen3-VL-Embedding performs best across all metrics and cutoffs, yet achieves only 49.23\% nDCG@10 and retrieves 66.92\% of the judged L1 products within the top 20. Its $\mathrm{R}_{L1}$ increases from 53.98\% at $K=10$ to 79.61\% at $K=50$, indicating that many optimal products appear only at deeper ranks. Moreover, its Hit$_{L1}$@1 is 47.02\%, meaning that an optimal product is ranked first for fewer than half of the queries. These results leave substantial room for improvement in both retrieving optimal products and ranking them near the top.

The results further reveal a clear gap between retrieving at least one optimal product and achieving broad category coverage. Qwen3-VL-Embedding retrieves at least one L1 product within the top 10 for 85.63\% of queries, yet its SC-R$_{L1}$@10 is only 61.33\% on the full query set and 55.86\% on $\mathcal{Q}_{\mathrm{multi}}$. This decline reflects the greater difficulty of covering multiple relevant leaf categories within a fixed retrieval budget. Overall, the substantial gap between Hit$_{L1}$@10 and SC-R$_{L1}$@10 suggests that existing models can often identify one optimal product but struggle to retrieve a set of products that covers the diverse solution categories associated with a query.

UME-R1 (Gen), which performs chain-of-thought reasoning before producing the query embedding, consistently outperforms UME-R1 (Disc) across all reported metrics. The gains, however, are small, with nDCG@10 improving by only 1.18 points and $\mathrm{R}_{L1}$@10 by 1.01 points. Within UME-R1, these results indicate consistent but limited benefits from reasoning before embedding generation, leaving a substantial gap to the strongest embedding retriever.

\subsection{Reranking Results}
\label{sec:reranking}

\begin{table}[t]
\centering
\caption{Reranking results on MIPR. All rerankers operate on the top-50 candidates retrieved by Qwen3-VL-Embedding-8B. The oracle reports the judgment-based upper bound achievable by reordering this candidate pool.}
\label{tab:reranking_results}
\small
\begin{tabular}{lcccc}
\toprule
Model & nDCG & $\mathrm{R}_{L1}$ & $\mathrm{Hit}_{L1}$ & $\mathrm{SC\text{-}R}_{L1}$ \\
\cmidrule(lr){2-2}
\cmidrule(lr){3-3}
\cmidrule(lr){4-4}
\cmidrule(lr){5-5}
& @10 & @10 & @1 & @10 ($\mathcal{Q}_{\mathrm{multi}}$) \\
\midrule
Embedding Baseline & 49.23 & 53.98 & 47.02 & 55.86 \\
\midrule
Lychee-Rerank-MM & \underline{55.85} & \underline{62.37} & \underline{53.80} & \underline{63.57} \\
Qwen2-VL-7B-Instruct & 46.72 & 53.07 & 43.02 & 56.21 \\
Qwen3-VL-8B-Instruct & 54.78 & 59.82 & 53.29 & 62.99 \\
Qwen3-VL-Reranker-8B & \textbf{60.11} & \textbf{66.00} & \textbf{59.26} & \textbf{67.00} \\
\midrule
Oracle & 86.04 & 78.95 & 94.97 & 80.22 \\
\bottomrule
\end{tabular}
\end{table}

We further evaluate whether rerankers can better exploit the retrieved candidates to improve both item-level relevance and category coverage. All rerankers operate on the top-50 candidates returned by Qwen3-VL-Embedding-8B and are evaluated using nDCG@10, $\mathrm{R}_{L1}$@10, $\mathrm{Hit}_{L1}$@1, and $\mathrm{SC\text{-}R}_{L1}$@10 on $\mathcal{Q}_{\mathrm{multi}}$. We also report a judgment-based candidate-pool oracle, which first ranks one L1 candidate from each distinct L1 leaf category represented in the top-50 candidate pool, followed by the remaining L1 candidates, L2 candidates, and all other candidates.

Table~\ref{tab:reranking_results} shows that Qwen3-VL-Reranker-8B performs best across all metrics. Compared with the embedding baseline, it improves nDCG@10 by 10.88 points, $\mathrm{R}_{L1}$@10 by 12.02 points, and $\mathrm{Hit}_{L1}$@1 by 12.24 points. It also raises $\mathrm{SC\text{-}R}_{L1}$@10 on $\mathcal{Q}_{\mathrm{multi}}$ from 55.86\% to 67.00\%. These results suggest that the initial retrieval stage identifies many useful candidates but does not rank them effectively. Notably, both specialized rerankers outperform the general-purpose MLLMs across all metrics, suggesting the benefit of reranking-specific training. Moreover, the best reranker remains 25.93, 12.95, 35.71, and 13.22 points below the candidate-pool oracle in nDCG@10, $\mathrm{R}_{L1}$@10, $\mathrm{Hit}_{L1}$@1, and $\mathrm{SC\text{-}R}_{L1}$@10, respectively. This gap indicates substantial room for better ranking the candidates already retrieved.

\section{Analysis and Discussion}
\label{sec:analysis}

In this section, we provide further analysis and discussion of MIPR. We conduct our analyses using five representative embedding retrievers from four model families: Qwen3-VL-Embedding, GME-Qwen2VL, LamRA-Ret, and UME-R1 in both its discriminative and generative variants.

\subsection{Modality Ablation and Intent Explicitness}
\label{sec:analysis_modalities}

\begin{figure}[t]
  \centering
 \includegraphics[width=\linewidth]{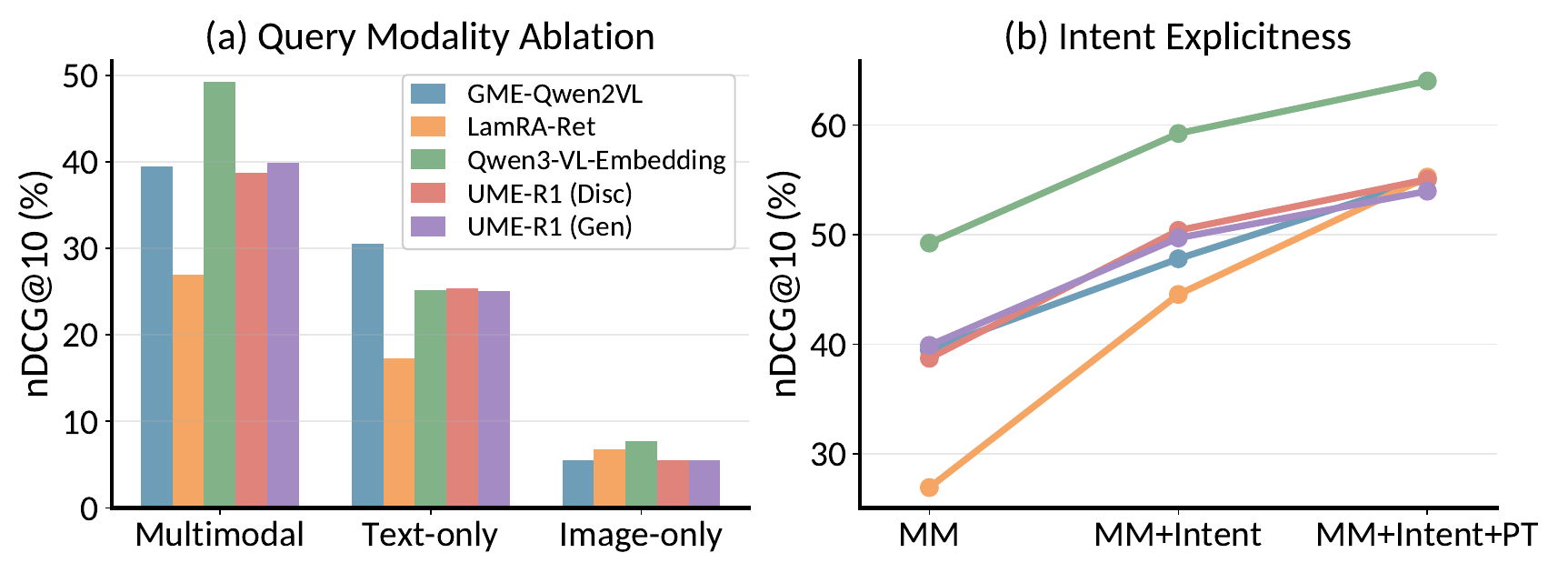}
  \caption{nDCG@10 under modality ablation and query augmentation. (a) Performance using image-only, text-only, and multimodal queries. (b) Performance after augmenting multimodal queries with intent descriptions and generated product terms.}
  \Description{Performance of five embedding-model variants under different query inputs. Multimodal queries outperform image-only and text-only queries, while adding intent descriptions and generated product terms further improves performance.}
\label{fig:modalities}
\end{figure}

MIPR queries combine visual context with a situational question. To examine the contribution of each modality, we evaluate representative retrievers using image-only, text-only, and multimodal queries, while keeping the product corpus and product representations unchanged. As shown in Figure~\ref{fig:modalities}(a), multimodal queries consistently outperform both single-modality variants across all models. Averaged across retrievers, nDCG@10 increases from 6.20\% with image-only queries and 24.70\% with text-only queries to 38.84\% with multimodal queries. The question alone lacks the object-specific context provided by the image, while the image alone does not express the user's need or concern. These results demonstrate that both modalities provide essential information for retrieval on MIPR.

To examine the role of intent inference, we further augment each multimodal query first with its intent description and then additionally with the generated product terms from our construction pipeline, making the connection between the query and suitable products increasingly explicit. As shown in Figure~\ref{fig:modalities}(b), Qwen3-VL-Embedding improves from 49.23\% nDCG@10 with the original query to 59.20\% after adding the intent description, and further reaches 64.00\% after additionally adding the generated product terms. The other models exhibit the same trend. These construction-informed augmentations provide a diagnostic reference rather than a controlled estimate of intent-inference difficulty. Nevertheless, the consistent gains suggest that explicit intent and solution cues can substantially ease retrieval on MIPR.

\begin{figure*}[t]
  \centering
 \includegraphics[width=\linewidth]{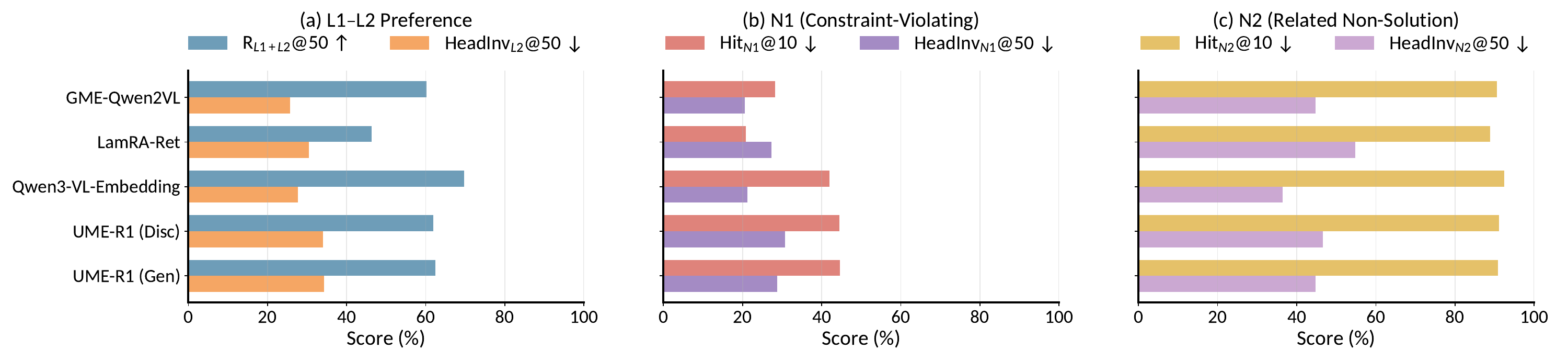}
  \caption{Retrieval and ordering results for graded solutions and hard negatives. (a) $\mathrm{R}_{L1+L2}$@50 and HeadInv$_{L2}$@50 for optimal and suboptimal products. (b) Hit$_{N1}$@10 and HeadInv$_{N1}$@50 for constraint-violating products. (c) Hit$_{N2}$@10 and HeadInv$_{N2}$@50 for related non-solutions.}
  \Description{Three bar charts comparing embedding retrievers. The first reports recall over optimal and suboptimal products and how often a suboptimal product precedes an optimal product. The second and third report hit rates and ordering inversions for constraint-violating products and related non-solutions, respectively.}
\label{fig:orderpref_hardneg}
\end{figure*}

\subsection{Preference for Optimal over Suboptimal Solutions}
\label{sec:analysis_graded}

We further analyze whether retrievers prioritize optimal (L1) solutions over suboptimal (L2) ones. We introduce \textbf{HeadInv$_{L2}$@$K$}, which measures how often the highest-ranked L2 product precedes the highest-ranked L1 product:
\begin{equation}
\small
\mathrm{HeadInv}_{L2}@K
=
\frac{1}{|\mathcal{Q}_{L1\cap L2}^{K}|}
\sum_{q\in\mathcal{Q}_{L1\cap L2}^{K}}
\mathbbm{1}\!\left[
\min_{j\in L_2(q)}\mathrm{rank}(j)
<
\min_{i\in L_1(q)}\mathrm{rank}(i)
\right].
\end{equation}
Here, $\mathcal{Q}_{L1\cap L2}^{K}$ denotes the set of queries whose top-$K$ results contain at least one L1 and one L2 product. $L_1(q)$ and $L_2(q)$ denote the retrieved L1 and L2 products for query $q$, respectively. Lower values indicate a stronger preference for optimal solutions. We additionally report $\mathrm{R}_{L1+L2}$@$K$, the macro-averaged fraction of judged L1 and L2 products retrieved within the top $K$. Because $\mathcal{Q}_{L1\cap L2}^{K}$ is model-dependent, we report the sizes and full results in Appendix~\ref{app:l1_l2_preference}.

As shown in Figure~\ref{fig:orderpref_hardneg}(a), all models exhibit an overall preference for L1 over L2 products, with HeadInv$_{L2}$@50 below 50\%. Nevertheless, the values range from 25.7\% to 34.4\%. Thus, among queries for which both L1 and L2 products appear in the top 50, an L2 product precedes every retrieved L1 product in more than one-quarter of cases. Retrieval and ordering performance are also not fully aligned. The two UME-R1 variants achieve $\mathrm{R}_{L1+L2}$@50 scores between 61.9\% and 62.5\%, compared with 60.2\% for GME-Qwen2VL. On their respective eligible query sets, however, the two UME-R1 variants have HeadInv$_{L2}$@50 values of approximately 34\%, substantially higher than the 25.7\% of GME-Qwen2VL. These results suggest that retrieving valid solutions and correctly prioritizing optimal ones remain distinct challenges.

\subsection{Hard-Negative Analysis}
\label{sec:analysis_hardneg}

We analyze model behavior on two types of hard negatives: constraint-violating (N1) products that appear applicable but violate critical user constraints, and related non-solution (N2) products that are related to the query but do not fulfill the user's intent. \textbf{Hit$_{N1}$@10} and \textbf{Hit$_{N2}$@10} measure the proportion of queries with at least one judged negative of the corresponding type for which such a negative appears in the top 10. Following Section~\ref{sec:analysis_graded}, we further report \textbf{HeadInv$_{N1}$@50} and \textbf{HeadInv$_{N2}$@50}. Each metric measures how often the highest-ranked negative precedes the highest-ranked L1 product among queries whose top-50 results contain both the corresponding negative type and an L1 product. Lower values are better for all four metrics. The eligible query sets for Hit$_{N1}$@10 and Hit$_{N2}$@10 are fixed across models, whereas those for the HeadInv metrics are model-dependent. Because negative hit rates are also affected by a model's overall retrieval ability, we interpret them together with the corresponding HeadInv metrics. More details and results are provided in Appendix~\ref{app:hard_negative}.

Figure~\ref{fig:orderpref_hardneg}(b) shows that constraint-violating products remain difficult to exclude and correctly rank. Hit$_{N1}$@10 ranges from 20.9\% to 44.7\% across models. GME-Qwen2VL achieves the lowest HeadInv$_{N1}$@50 at 20.6\%, closely followed by Qwen3-VL-Embedding at 21.3\%. However, Qwen3-VL-Embedding has a higher Hit$_{N1}$@10 than GME-Qwen2VL (42.0\% vs. 28.3\%). UME-R1 (Gen) reduces HeadInv$_{N1}$@50 from 30.8\% to 28.8\% over its discriminative variant, but still trails the two strongest models. These results show that retrieving N1 products for a larger proportion of eligible queries does not necessarily imply a weaker ability to rank L1 products ahead of them.

As shown in Figure~\ref{fig:orderpref_hardneg}(c), related non-solutions appear much more frequently near the top of the ranking. Hit$_{N2}$@10 ranges from 88.9\% to 92.5\%, indicating that all models frequently retrieve N2 products. Qwen3-VL-Embedding has the highest Hit$_{N2}$@10 at 92.5\%, but achieves the lowest HeadInv$_{N2}$@50 at 36.5\%. It therefore retrieves N2 products for many eligible queries while frequently ranking an L1 product ahead of them when both types occur in the top 50. In comparison, GME-Qwen2VL and LamRA-Ret reach HeadInv$_{N2}$@50 values of 44.8\% and 54.8\%, respectively. HeadInv$_{N2}$@50 is also consistently higher than HeadInv$_{N1}$@50 across models. Although the two metrics are computed over different eligible query sets, this pattern suggests that distinguishing optimal products from related non-solutions may be more difficult than distinguishing them from constraint-violating products.

\subsection{Performance by Number of Solution Categories}
\label{sec:analysis_solution_number}

\begin{figure}[t]
  \centering
 \includegraphics[width=\linewidth]{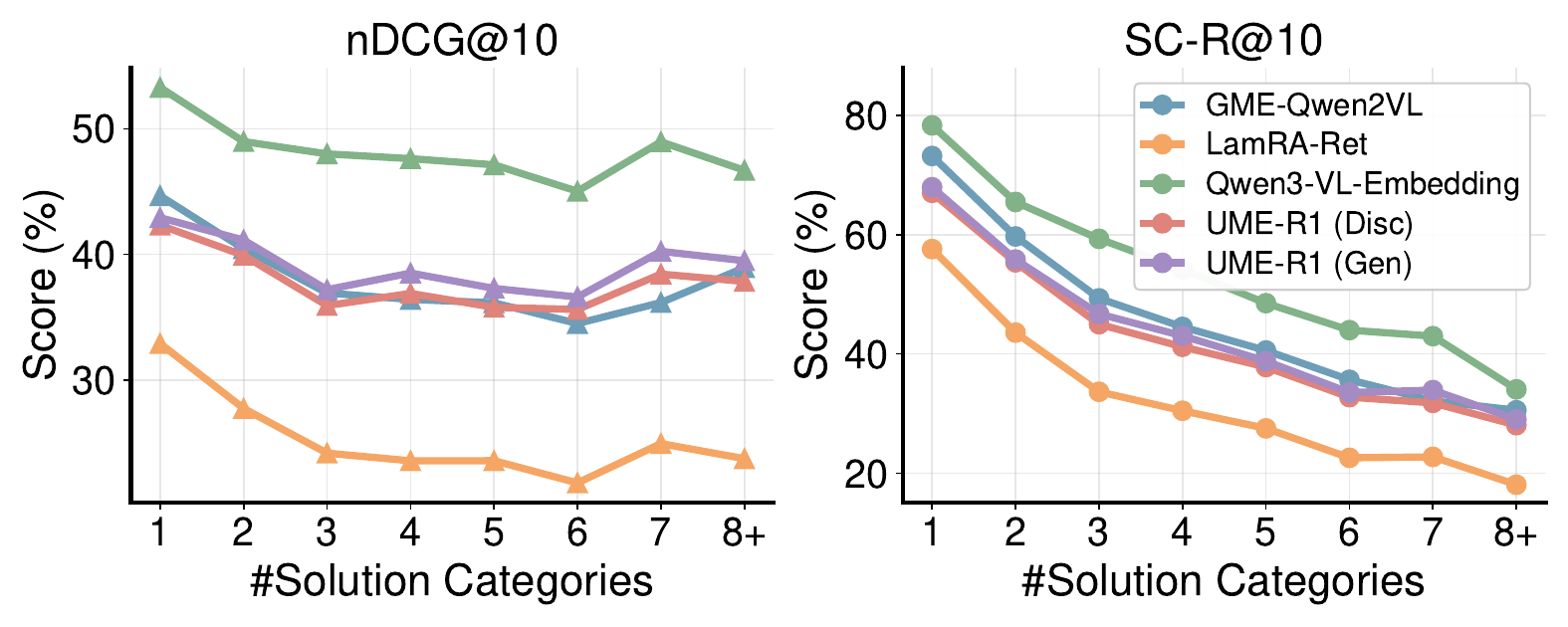}
  \caption{nDCG@10 and $\mathrm{SC\text{-}R}_{L1}$@10 for queries grouped by the number of annotated L1 solution categories $|\mathcal{S}(q)|$.}
  \Description{Two line charts showing nDCG at 10 and L1 solution-category recall at 10 for embedding retrievers as the number of annotated L1 solution categories increases. nDCG decreases moderately, while solution-category recall declines substantially.}
\label{fig:solution_number}
\end{figure}

To examine how model performance varies with the number of solution categories, we group queries by the number of annotated L1 solution categories, ranging from 1 to $8{+}$, and report nDCG@10 and $\mathrm{SC\text{-}R}_{L1}$@10 for each group in Figure~\ref{fig:solution_number}. The two metrics increasingly diverge as the number of solution categories grows. For Qwen3-VL-Embedding, nDCG@10 decreases moderately from 53.3\% to 46.7\%, indicating that it remains relatively effective at ranking some relevant products near the top. In contrast, $\mathrm{SC\text{-}R}_{L1}$@10 drops from 78.3\% to 34.1\%, with a similar trend across models. This divergence shows that ranking relevant products near the top does not ensure broad coverage of the annotated solution space. As the number of solution categories grows, covering them within a fixed top-10 ranking becomes increasingly difficult, leaving many annotated categories uncovered.

\subsection{Impact of Pooling-Based Judgment Expansion}
\label{sec:analysis_pooling}

We examine the impact of pooling-based judgment expansion by comparing evaluation scores computed using the initial and final qrels. Averaged across the ten embedding retrievers, expansion increases nDCG@10 and Hit$_{L1}$@1 by 3.02 and 2.77 points, respectively, as additional judgments allow previously unjudged relevant products to receive credit. In contrast, $\mathrm{R}_{L1}$@10 decreases by 1.45 points on average because expansion also enlarges the set of judged L1 products in the recall denominator. Despite these changes in absolute scores, model rankings remain highly stable, with Kendall's $\tau$ ranging from 0.911 to 1.000 across metrics; the few rank reversals occur only between models with similar performance. These results indicate that judgment expansion improves evaluation coverage while preserving the main comparative conclusions. Full results are reported in Appendix~\ref{app:pooling_impact}.

\section{Conclusion}
\label{sec:conclusion}
We present MIPR, a multimodal benchmark for intent-to-product retrieval constructed from real Chinese e-commerce search logs. MIPR jointly captures reasoning-intensive retrieval, multi-category solutions, and fine-grained relevance labels. Experiments show that embedding retrievers often retrieve useful candidates but struggle to rank optimal products highly and cover multiple solution categories. Although reranking substantially improves performance, the large gap to the candidate-pool oracle indicates considerable room for improvement. We hope MIPR and its construction pipeline will facilitate future research on this challenging retrieval setting.

\bibliographystyle{ACM-Reference-Format}
\bibliography{sample-base}

\newpage
\appendix

\section{Ethics, Privacy, and Data Compliance}
\label{app:data_compliance}

MIPR is constructed from Taobao search logs and product listings collected and processed under the platform's user agreements, privacy policies, and applicable data-governance requirements. The data are used solely for research on multimodal retrieval. Query and product identifiers are replaced with hashed IDs, and personally identifiable information and user-identifying metadata are removed.

The benchmark data, annotations, and code are undergoing Taobao's internal privacy, security, copyright, and compliance review, which will be completed prior to release. This review covers both textual and visual content, including potential personal information and intellectual-property concerns. Upon acceptance, the processed queries, product data, relevance judgments, construction code, and evaluation code will be publicly released.

\section{Additional Construction Details and Pipeline Validation}
\label{app:pipeline_validation}

This section provides additional details on the human validation conducted at key stages of the construction pipeline. Two native Chinese speakers with prior experience in relevance annotation participated: a master’s student (Annotator A) performed the primary validation, while a doctoral student (Annotator B) independently reviewed randomly sampled subsets. Unless otherwise specified, samples were drawn independently for each validation study, and disagreements were resolved through discussion.

\subsection{Query Curation}
\label{app:query_validation}

\textbf{Eligibility criteria.}
MIPR retains queries for which the image and question jointly provide sufficient information to infer suitable products. We exclude queries with any of the following properties:
\begin{itemize}[nosep,leftmargin=*]
    \item \textbf{Subjective or similarity-seeking intent:} the query primarily asks for aesthetic coordination, subjective comparison, similar styles, or alternative colors, without sufficiently objective criteria for product relevance.
    \item \textbf{Insufficient product-selection information:} selecting a suitable product requires unavailable specifications, compatibility information, or price information that cannot be inferred from the image or question.
    \item \textbf{Incompatible task form:} the query requests a bundle or checklist of products that must be used together, directly specifies a target product, or can be resolved from the text alone without using the image.
    \item \textbf{Poor data quality:} the image is unrecognizable or clearly inconsistent with the question.
\end{itemize}

\textbf{Rule-based and lightweight filtering.}
We first apply regular-expression rules to remove queries matching predefined ineligible patterns, followed by conservative coarse filtering with Qwen3.5-9B. The rules and filtering prompt are included in the released code.

\textbf{Intent-analysis validation.}
Claude Sonnet 4.6 produces a structured analysis of each remaining query, including its visual context, user intent, and product-selection constraints. Annotator A inspected 200 randomly sampled retained queries and their corresponding structured analyses. Among these analyses, 96.5\% correctly captured both the relevant image content and the user intent. Of the seven analyses containing errors, only one contained an error substantial enough to alter the inferred intent or applicable constraints.

\textbf{Eligibility validation.}
GLM-5.2 and DeepSeek-V4-Pro~\cite{xu2026deepseekv4} independently assess each query based on the original image--question pair and its structured analysis, and only queries accepted by both models are retained. Annotator A inspects 200 randomly sampled retained queries, of which only 1.5\% are judged unsuitable or borderline. Annotator B independently reviewed 10 randomly selected cases from this sample and agreed with Annotator A’s judgments in all cases.

\textbf{Redundancy reduction.}
We encode the inferred intent of each retained query using Qwen3-Embedding-8B~\cite{zhang2025qwen3embedding} and apply non-transitive leader clustering at a cosine-similarity threshold of $\tau=0.70$. Queries are processed in descending order of centrality; each unassigned query becomes a leader and absorbs all unassigned queries directly similar to it, avoiding the chained clusters caused by transitive linkage. We retain the query closest to the cluster centroid as the representative. This step reduces 9.8K queries to 3.7K.

Annotator A inspects the 20 largest clusters and finds that each is dominated by a single recurring intent pattern. Among the resulting clusters, 2,366 are singletons, indicating that the reduction removes recurring patterns while retaining a diverse set of intents.

\subsection{Candidate Product Discovery}
\label{app:candidate_validation}

\textbf{Coverage of generated terms.}
Candidate discovery starts with a two-stage enumeration-and-reflection procedure using Qwen3.6-27B. To assess category coverage, Annotator A manually enumerated plausible product terms for 50 randomly sampled queries. For each term, we retrieved the top-5 products from the source catalog using the same text-embedding retrieval procedure as in the main pipeline and manually verified their validity. The leaf categories of the verified products were then merged to form a retrieval-grounded human reference set. The initial model enumeration covered 81\% of the categories in this reference set, with coverage increasing to 89\% after reflection. In addition, reflection identified an average of 0.74 manually verified valid categories per query that were absent from the human reference set, indicating that the manually constructed reference is not exhaustive.

\textbf{Complementary retrieval routes.}
We combine term-based retrieval with direct intent-based retrieval. The retrieval routes include BM25 with jieba tokenization, two text embedding models (BGE-small-zh-v1.5~\cite{xiao2024bge} and Qwen3-Embedding-0.6B~\cite{zhang2025qwen3embedding}) whose rankings are fused using reciprocal rank fusion (RRF), and one multimodal embedding model (Qwen3-VL-Embedding-2B~\cite{li2026qwen3-vl-embedding}). Each product term is issued to all three routes, and the inferred intent is additionally issued to the fused text embedding route and the multimodal embedding model. Every route retains its top-5 products, so a query with $n$ product terms yields at most $5(3n+2)$ candidates. After merging by product ID, each query retains 50 candidate products on average.

\subsection{Shared Corpus Construction}
\label{app:corpus_validation}

\textbf{Product deduplication.}
To reduce redundancy in the union of query-level candidates, we encode each product's image and title using Qwen3-VL-Embedding-8B~\cite{li2026qwen3-vl-embedding} and perform similarity-based clustering. Within each leaf category, we apply leader clustering with a cosine-similarity threshold of 0.60. This reduces the corpus from 53,786 to 30,731 products and produces 7,347 non-singleton clusters. We retain the product with the highest average similarity to the other members as the representative and redirect the associated query--candidate pairs accordingly.

\textbf{Manual inspection.}
As a sanity check, one annotator inspected the 20 largest non-singleton clusters. These clusters primarily contain near-duplicate or interchangeable listings of standardized products, such as screwdrivers, shoe cleaners, and insoles, often differing only in packaging or vendor. No inspected cluster contains functionally distinct products.

\subsection{Fine-Grained Relevance Annotation}
\label{app:screening_validation}

\textbf{Coarse screening.}
Qwen3.6-27B performs conservative binary screening designed to retain ambiguous, partially applicable, visually related, and potentially constraint-violating candidates while assigning N3 to clearly unrelated products. Manual inspection of all screened-out products for 50 randomly sampled queries yields a false-rejection rate of 4.4\%, where a false rejection is a product that should have received L1, L2, N1, or N2.

\textbf{Five-way annotation.}
The retained candidates are independently labeled by Qwen3.7-Max, GLM-5.2, and DeepSeek-V4-Pro. We use majority voting, with Claude Sonnet 4.6 adjudicating cases in which the model judgments cross the positive--negative boundary (\{L1, L2\} versus \{N1, N2, N3\}) or contain three distinct labels. The three models agree unanimously on 73.69\% of candidates, while 7.14\% require adjudication, including 3.18\% that receive three distinct initial labels. The remaining non-unanimous cases are resolved by majority vote. Their dissenting labels typically differ only in degree within the same polarity, most often between L1 and L2 or between N2 and N3.

\textbf{Validation against human annotations.}
Annotator A labeled all initially discovered candidates for 50 randomly sampled queries, including candidates assigned N3 during coarse screening, by directly inspecting the original queries, product images, and accompanying text. Annotator B independently labeled all candidates associated with ten randomly selected queries from this sample. Before reconciliation, the two annotators achieved 87.8\% exact agreement under the five-way schema, with most disagreements occurring at the N2--N3 boundary. Disagreements in the double-annotated subset were resolved through discussion, while Annotator A's labels were used for the remaining queries, together forming the human reference. Against this reference, the complete annotation pipeline achieved 81.3\% exact agreement and 81.4\% macro-F1. The per-label recalls for L1, L2, N1, N2, and N3 were 83.5\%, 81.8\%, 85.4\%, 76.9\%, and 74.5\%, respectively, with corresponding precisions of 85.8\%, 81.7\%, 88.2\%, 79.9\%, and 76.4\%. The relatively lower performance on N2 and N3 is consistent with the ambiguity observed at this boundary during human annotation.

\subsection{Pooling-Based Judgment Expansion}
\label{app:pooling}

\paragraph{Pool construction.}
Judgment pools are constructed using the ten multimodal embedding models evaluated in Section~\ref{sec:experimental_setup}, with each model contributing its top-20 products per query. Using the evaluated models as pooling systems ensures direct judgment coverage of the ranked lists reported in our experiments. The models also differ in architecture and training data, providing complementary candidate sets. We merge their retrieved products by product ID and annotate every previously unjudged query--product pair using the same coarse-screening and five-way annotation procedure described in Section~\ref{sec:relevance_annotation}.

\paragraph{Pool statistics.}
Pooling adds 299,479 judgments, increasing the average number of judged products per query from 42.1 to 127.1. The additional judgments contain, on average, 1.44 L1, 3.70 L2, 1.77 N1, 15.11 N2, and 63.00 N3 products per query. In addition to identifying previously unjudged positive products, pooling therefore provides broader coverage of challenging negatives, particularly constraint-violating (N1) and related non-solution (N2) products.

\paragraph{Coverage and limitations.}
Because the pools are constructed from a fixed set of evaluated models, future systems may retrieve relevant products that remain unjudged and may therefore be affected by incomplete-judgment bias. As in standard pooled evaluation, pairs outside the judgment pools are treated as non-relevant. We release the complete pool memberships, per-model contributions, and final qrels to enable future judgment expansion.

\section{Additional Experimental Details and Results}
\label{app:additional_analysis}

\subsection{Reranker Implementation}
\label{app:reranker_details}

All rerankers use the following prompt:
\begin{promptbox}
\small

{\color{promptlabel}\bfseries System}\\
Judge whether the Document meets the requirements based on the Query and the
Instruct provided. Note that the answer can only be ``yes'' or ``no.''

\medskip
{\color{promptlabel}\bfseries User}\\
\texttt{<Instruct>:} Given a search query containing a reference product image
and the user's need, retrieve relevant products that can satisfy the user's need.

\smallskip
\texttt{<Query>:}
\fcolorbox{promptborder}{white}{\texttt{query\_image}}
\ \texttt{\{query\_text\}}

\smallskip
\texttt{<Document>:}
\fcolorbox{promptborder}{white}{\texttt{product\_image}}
\ \texttt{\{product\_title\}}
\end{promptbox}

For Qwen2-VL-7B-Instruct, Qwen3-VL-8B-Instruct, and Lychee-Rerank-MM, we extract the next-token logits for \texttt{yes} and \texttt{no} and compute the relevance score as
\begin{equation}
s(q,p)=
\frac{\exp(z_{\texttt{yes}})}
{\exp(z_{\texttt{yes}})+\exp(z_{\texttt{no}})}.
\end{equation}
For Qwen3-VL-Reranker-8B, we use the equivalent memory-efficient implementation provided by its official code. Given the final hidden state $h$, the score is
\begin{equation}
s(q,p)=
\sigma\left((w_{\texttt{yes}}-w_{\texttt{no}})^\top h\right)
=\sigma(z_{\texttt{yes}}-z_{\texttt{no}}),
\end{equation}
where $w_{\texttt{yes}}$ and $w_{\texttt{no}}$ are the corresponding output-head vectors. This is mathematically equivalent to the binary softmax above but avoids computing logits over the full vocabulary. Candidates are ranked in descending order of their relevance scores.

\subsection{Additional Results for L1--L2 Preference}
\label{app:l1_l2_preference}

Table~\ref{tab:l1_l2_preference_full} reports additional results for the analysis in Section~\ref{sec:analysis_graded}. $\mathrm{R}_{L1+L2}$@50 is the macro-averaged recall over the union of judged L1 and L2 products. It is computed over all 3,522 queries and uses the same top-50 cutoff as the preference analysis.

HeadInv$_{L2}$@50 is computed over $\mathcal{Q}_{L1\cap L2}^{50}$, the set of queries for which the retrieved top 50 contains at least one L1 product and one L2 product. The column $n$ reports the model-dependent size of this set. Because the eligible set differs across models, HeadInv$_{L2}$@50 should be interpreted as a conditional ordering diagnostic: it measures whether a model ranks an L1 product ahead of the highest-ranked L2 product when both solution levels are retrieved.

\begin{table}[t]
\centering
\caption{Additional L1--L2 retrieval and ordering results at cutoff 50.}
\label{tab:l1_l2_preference_full}
\small
\begin{tabular}{lccc}
\toprule
Model
& $\mathrm{R}_{L1+L2}$@50 $\uparrow$
& HeadInv$_{L2}$@50 $\downarrow$
& $n$ \\
\midrule
GME-Qwen2VL        & 60.19 & 25.72 & 2,228 \\
LamRA-Ret          & 46.31 & 30.49 & 1,771 \\
Qwen3-VL-Embedding & 69.77 & 27.69 & 2,510 \\
UME-R1 (Disc)      & 61.86 & 34.07 & 2,363 \\
UME-R1 (Gen)       & 62.50 & 34.37 & 2,371 \\
\bottomrule
\end{tabular}
\end{table}

\subsection{Additional Results for Hard-Negative Analysis}
\label{app:hard_negative}

Table~\ref{tab:hard_negative_full} provides additional results for the hard-negative analysis in Section~\ref{sec:analysis_hardneg}. Hit$_{N1}$@10 is computed over the 1,601 queries judged to contain at least one N1 product, while Hit$_{N2}$@10 is computed over the 3,521 queries judged to contain at least one N2 product. Their denominators are therefore fixed across models.

HeadInv$_{N1}$@50 and HeadInv$_{N2}$@50 are computed over model-dependent query sets. Specifically, HeadInv$_{Nt}$@50 for $Nt\in\{N1,N2\}$ is evaluated only on queries whose top-50 results contain at least one L1 product and one product of the corresponding negative type. The columns $n_{N1}$ and $n_{N2}$ report the sizes of these query sets.

The hit and HeadInv metrics capture different aspects of retrieval. A lower negative hit rate indicates fewer judged hard negatives in the top 10, but may also result from weaker overall retrieval. In contrast, HeadInv measures whether a model ranks an L1 product ahead of a retrieved hard negative, conditional on both appearing in the top 50. The two metrics should therefore be interpreted jointly.

\begin{figure}[t]
    \centering
    \includegraphics[width=1\linewidth]{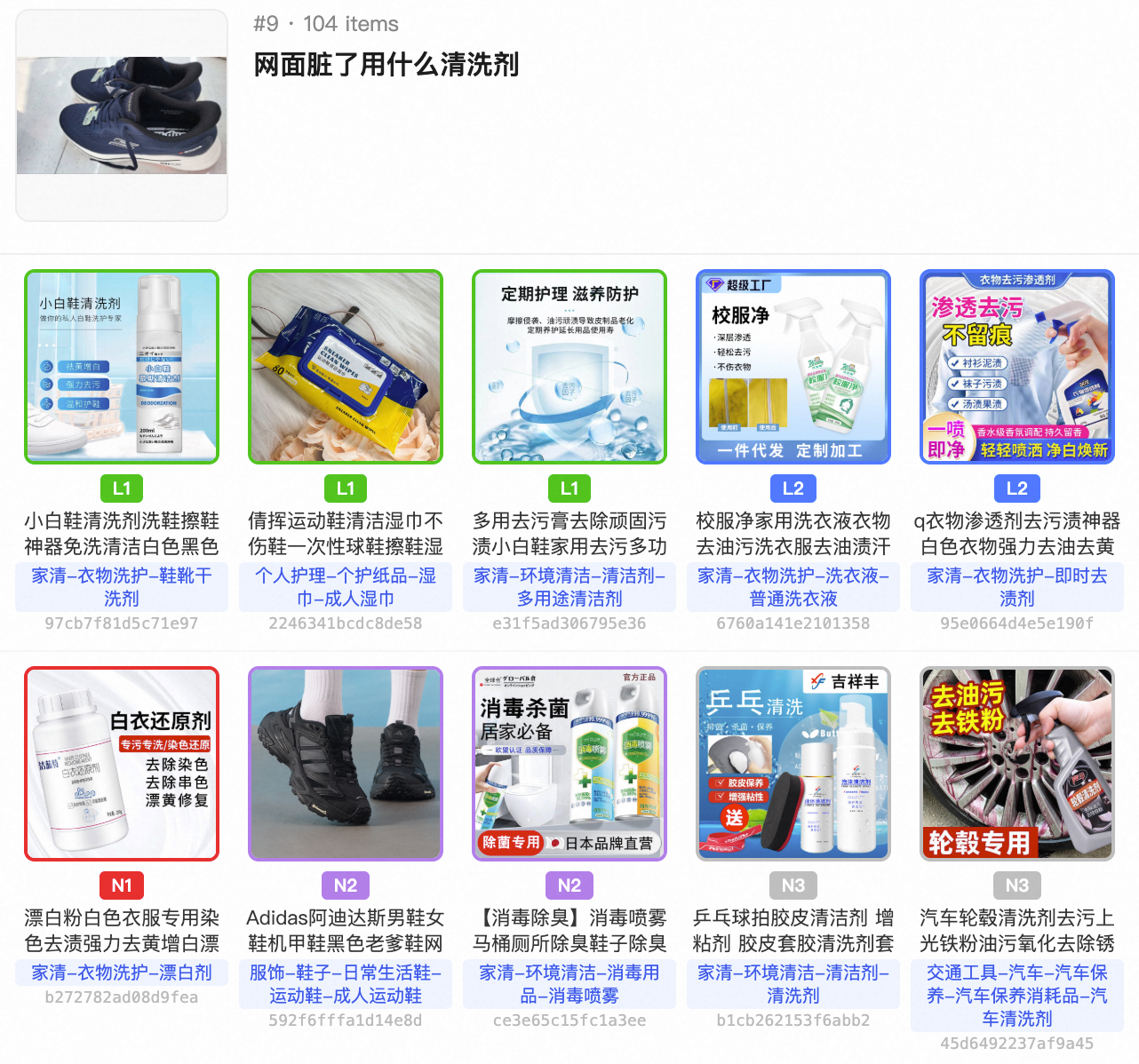}
    \includegraphics[width=1\linewidth]{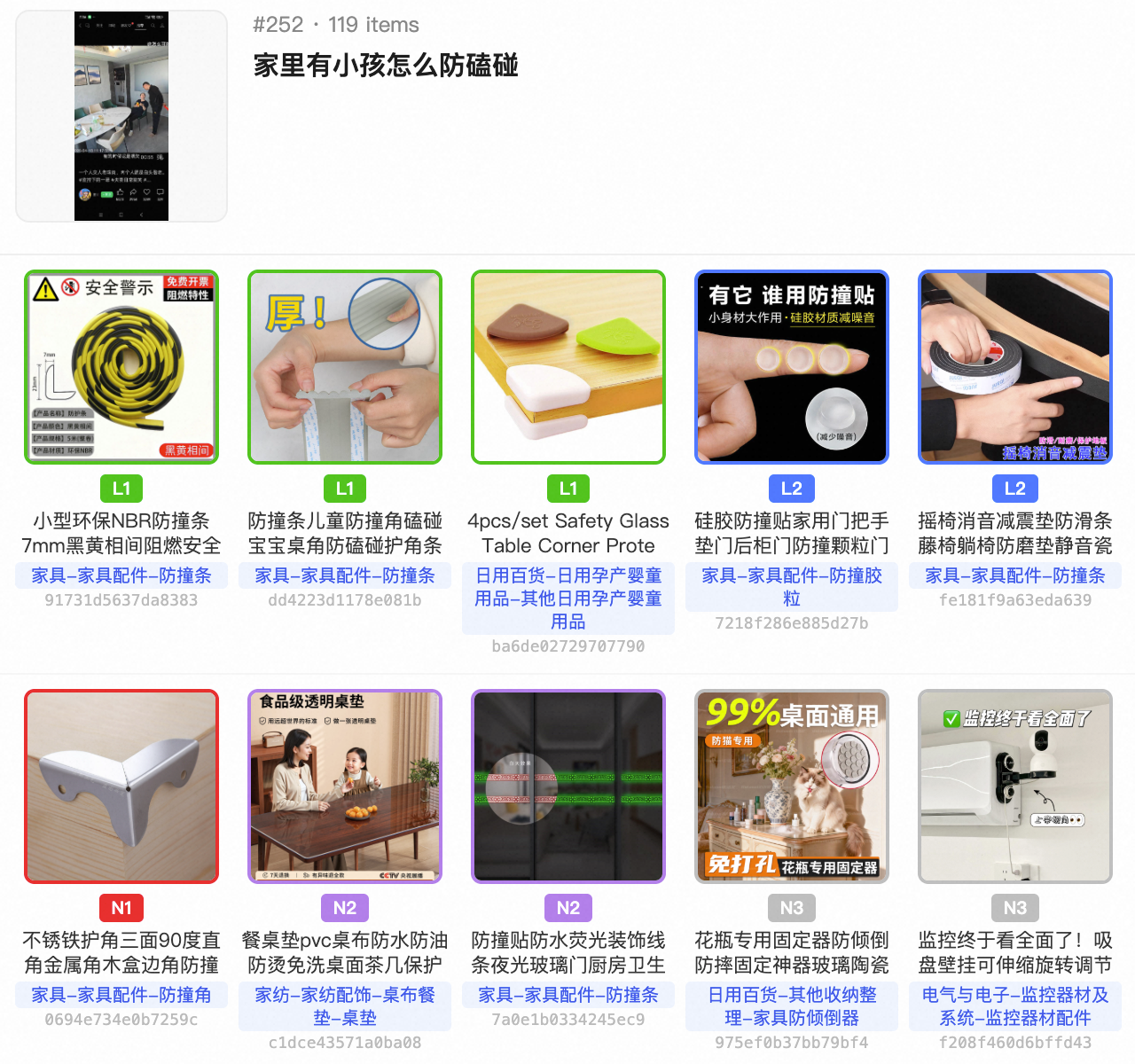}
    \caption{Two representative examples from MIPR.}
    \Description{Two multimodal queries with representative products covering all five relevance labels: optimal, suboptimal, constraint-violating, related non-solution, and irrelevant.}
    \label{fig:dataset_case}
\end{figure}

\subsection{Additional Results for Performance by Number of Solution Categories}
\label{app:solution_category_distribution}

\begin{table}[t]
\centering
\caption{Distribution of queries by the number of distinct judged L1 leaf categories.}
\label{tab:solution_category_distribution}
\small
\begin{tabular}{@{}crcr@{}}
\toprule
\multicolumn{1}{c}{L1 categories} & 
\multicolumn{1}{c}{Queries} &
\multicolumn{1}{c}{L1 categories} & 
\multicolumn{1}{c}{Queries} \\
\midrule
1 & 858 & 5      & 282 \\
2 & 840 & 6      & 172 \\
3 & 638 & 7      & 120 \\
4 & 434 & \(8+\) & 178 \\
\midrule
\multicolumn{3}{r}{Total} & 3,522 \\
\bottomrule
\end{tabular}
\end{table}

Table~\ref{tab:solution_category_distribution} reports the distribution of queries by the number of distinct judged L1 leaf categories. This grouping is used in the analysis in Section~\ref{sec:analysis_solution_number}. We retain separate groups for one through seven categories and merge queries with eight or more categories into a single $8+$ group. The latter contains queries with between 8 and 16 distinct L1 leaf categories.

\subsection{Additional Results for Impact of Pooling-Based Judgment Expansion}
\label{app:pooling_impact}

Table~\ref{tab:pooling_impact} reports the per-model effects of replacing the initial qrels with the final qrels after pooling-based judgment expansion. Expansion consistently increases nDCG@10, Hit$_{L1}$@1, and SC-R$_{L1}$@10, while $\mathrm{R}_{L1}$@10 decreases slightly because the set of judged L1 products grows. Model rankings remain stable across all four metrics. The few reversals occur only between models with similar scores.

\section{Dataset Examples}
\label{app:dataset_examples}

Due to space constraints, we present two representative examples from MIPR. As shown in Figure~\ref{fig:dataset_case}, each example includes a multimodal query and representative products covering all five relevance labels: L1 (optimal), L2 (suboptimal), N1 (constraint-violating), N2 (related non-solution), and N3 (irrelevant). Together, these examples illustrate the multi-category nature of valid solutions and the fine-grained distinctions captured by the relevance schema.

\begin{table*}[t]
\centering
\caption{Additional hard-negative retrieval and ordering results.}
\label{tab:hard_negative_full}
\begin{tabular}{lcccccc}
\toprule
Model
& Hit$_{N1}$@10 $\downarrow$
& Hit$_{N2}$@10 $\downarrow$
& HeadInv$_{N1}$@50 $\downarrow$
& HeadInv$_{N2}$@50 $\downarrow$
& $n_{N1}$
& $n_{N2}$ \\
\midrule
GME-Qwen2VL        & 28.29 & 90.63 & 20.60 & 44.78 & 893   & 3,209 \\
LamRA-Ret          & 20.86 & 88.90 & 27.34 & 54.83 & 662   & 2,865 \\
Qwen3-VL-Embedding & 41.97 & 92.45 & 21.29 & 36.46 & 1,132 & 3,327 \\
UME-R1 (Disc)      & 44.53 & 91.14 & 30.77 & 46.54 & 1,105 & 3,212 \\
UME-R1 (Gen)       & 44.66 & 90.88 & 28.79 & 44.77 & 1,108 & 3,201 \\
\bottomrule
\end{tabular}
\end{table*}

\begin{table*}[t]
\centering
\caption{Effect of pooling-based judgment expansion. Values outside parentheses are scores using the final qrels, while parentheses report changes relative to the initial qrels, i.e., $\Delta=\mathrm{final}-\mathrm{initial}$. The final row reports Kendall's $\tau$ between model rankings under the initial and final qrels.}
\label{tab:pooling_impact}
\setlength{\tabcolsep}{10pt}
\begin{tabular}{lcccc}
\toprule
Model & nDCG@10 & $\mathrm{R}_{L1}$@10 & $\mathrm{Hit}_{L1}$@1 & $\mathrm{SC\text{-}R}_{L1}$@10 ($\mathcal{Q}_{\mathrm{multi}}$) \\
\midrule
CORAL                    & 10.53 $(+1.05)$ & 12.94 $(-0.43)$ &  9.11 $(+1.06)$ & 13.13 $(+1.19)$ \\
GME-Qwen2VL           & 39.48 $(+2.57)$ & 46.78 $(-2.28)$ & 37.68 $(+3.18)$ & 47.98 $(+0.43)$ \\
LamRA-Ret                & 26.91 $(+1.33)$ & 32.94 $(-2.12)$ & 24.47 $(+1.79)$ & 33.36 $(+0.38)$ \\
LLaVE                 & 34.52 $(+5.50)$ & 36.30 $(-1.47)$ & 27.91 $(+3.60)$ & 39.87 $(+1.30)$ \\
MM-Embed                 & 27.84 $(+3.31)$ & 29.06 $(-2.45)$ & 24.56 $(+2.52)$ & 33.13 $(+0.16)$ \\
Qwen3-VL-Embedding    & 49.23 $(+4.84)$ & 53.98 $(-1.94)$ & 47.02 $(+4.70)$ & 55.86 $(+0.81)$ \\
UME-R1 (Disc)           & 38.71 $(+4.89)$ & 42.37 $(-1.53)$ & 34.78 $(+4.05)$ & 44.36 $(+0.96)$ \\
UME-R1 (Gen)         & 39.89 $(+5.37)$ & 43.38 $(-1.15)$ & 36.26 $(+4.83)$ & 45.54 $(+1.13)$ \\
VLM2Vec-Qwen2VL       & 13.80 $(+0.79)$ & 17.49 $(-0.87)$ & 11.75 $(+1.06)$ & 17.38 $(+0.58)$ \\
VLM2Vec-V2               &  7.51 $(+0.52)$ & 10.23 $(-0.28)$ &  6.50 $(+0.87)$ &  9.10 $(+0.33)$ \\
\midrule
Average                  & 28.84 $(+3.02)$ & 32.55 $(-1.45)$ & 26.00 $(+2.77)$ & 34.56 $(+0.73)$ \\
Kendall's $\tau$         & 0.911            & 1.000            & 0.956            & 1.000            \\
\bottomrule
\end{tabular}
\end{table*}

\section{Model-Assisted Construction Prompts}
\label{app:prompts}

This section presents abridged versions of the Chinese prompts used in the model-assisted stages of benchmark construction, including query filtering, candidate product-term generation, and fine-grained relevance annotation. To improve readability, we omit repeated instructions and few-shot examples while preserving the core decision criteria, input fields, and output formats. The complete prompts, including all examples and implementation-specific formatting, will be released with the code upon acceptance. Placeholders enclosed in braces denote fields populated automatically by the construction pipeline.

\subsection{Query Eligibility Prompt}

Below is an abridged version of the Chinese query-eligibility prompt. We omit
repeated explanations and few-shot examples for space. The complete prompt
will be released with the code.

\begin{CJK*}{UTF8}{gbsn}
\begin{promptbox}
\small

{\color{promptlabel}\bfseries System}\\
用户上传了一张图片并提出问题，希望购买商品来解决需求。请判断该 query 是否适合“图像与意图到商品检索”任务。

\medskip
{\color{promptlabel}\bfseries User}\\
适合的 query 应满足以下条件：图片提供商品选择所需的关键信息；用户表达了能够通过商品解决的明确需求；合适的商品可依据相对客观的标准确定。

判断时以图片内容和用户原始问题为主要依据，推断出的用户意图仅作为辅助。如果用户意图与图片或原始问题矛盾，以后两者为准。

若 query 符合以下任一条件，则将其丢弃：

\begin{itemize}[nosep,leftmargin=*]
    \item \textbf{A1（搭配或审美）：}答案主要取决于个人审美、风格或穿搭偏好。
    \item \textbf{A2（主观比较或优选）：}需要进行缺乏客观标准的产品比较或偏好权衡。
    \item \textbf{A3（相似款或配色）：}目标是寻找同款、相似款式或其他配色。
    \item \textbf{B1（缺少规格）：}商品选择依赖图片和文本中均无法确定的精确尺寸、容量或型号。
    \item \textbf{B2（缺少配件参数）：}配件选择依赖不可获得的详细兼容性参数。
    \item \textbf{B3（价格约束）：}需求包含当前商品库无法验证的价格限制。
    \item \textbf{C1（清单或成套需求）：}用户同时需要多个不同类型的商品组成清单或套装。
    \item \textbf{C2（纯文本可解）：}仅根据文本即可确定商品，图片未提供必要信息。
    \item \textbf{D1（数据质量）：}图片无法识别，或图片与问题明显不匹配。
\end{itemize}

若没有命中任何丢弃条件，则保留该 query。若多个条件同时适用，输出最主要的一个原因。

\medskip
\textbf{输出}\\
仅输出一个合法的 JSON 对象，不要输出其他内容：

\smallskip
保留：\\
\texttt{\{"decision":"keep","discard\_reason":null\}}

\smallskip
丢弃：\\
\texttt{\{"decision":"discard","discard\_reason":"CODE"\}}

其中 \texttt{CODE} 为上述丢弃条件对应的编号。
\end{promptbox}
\end{CJK*}

\subsection{Candidate Product-Term Generation Prompts}

Product terms are generated through two stages: initial enumeration and reflection. Below are abridged versions of the Chinese prompts; repeated instructions and few-shot examples are omitted for space. The complete prompts will be released with the code.

Initial enumeration.

\begin{CJK*}{UTF8}{gbsn}
\begin{promptbox}
\small

{\color{promptlabel}\bfseries System}\\
用户上传了一张图片并提出问题，希望购买商品来解决需求。请生成可用于淘宝商品检索的关键词。

\medskip
{\color{promptlabel}\bfseries User}\\
根据以下信息理解用户需求。其中，用户意图是主要依据，图片内容是事实锚点，原始问题和选品注意作为补充参考。

从以下四个角度考虑可能的商品：
\begin{itemize}[nosep,leftmargin=*]
    \item \textbf{专用品：}为该需求专门设计的直接解决方案；
    \item \textbf{通用替代品：}能够实现相同或相近效果的通用商品；
    \item \textbf{配套品：}解决需求所需的配件、耗材或替换件；
    \item \textbf{前置品：}实施解决方案前需要具备的商品。
\end{itemize}

并非每个角度都必须产生关键词。仅生成能够从图片和需求中合理推断的商品，不要凭空扩展。

关键词须满足以下要求：
\begin{itemize}[nosep,leftmargin=*]
    \item 必须对应淘宝可购买的实物商品；
    \item 覆盖不同商品类型或解决方式，不使用同义词重复凑数；
    \item 表述简洁、具体，符合真实电商搜索习惯；
    \item 至少生成一个、最多生成六个关键词；若合理方案较少，无需凑满。
\end{itemize}

\medskip
\textbf{输入}\\
用户意图：\texttt{\{intent\}}\\
图片内容：\texttt{\{content\}}\\
用户提问：\texttt{\{query\}}\\
选品注意：\texttt{\{note\}}

\medskip
\textbf{输出}\\
仅输出关键词组成的合法 JSON 列表，不要输出其他内容。

\end{promptbox}
\end{CJK*}

Reflection and refinement.

\begin{CJK*}{UTF8}{gbsn}
\begin{promptbox}
\small

{\color{promptlabel}\bfseries System}\\
用户上传了一张图片并提出问题，希望购买商品来解决需求。你已生成一组淘宝搜索关键词；请对其进行复查并输出最终列表。

\medskip
{\color{promptlabel}\bfseries User}\\
根据用户意图、图片内容、原始问题和选品注意，依次执行以下检查：

\begin{enumerate}[nosep,leftmargin=*]
    \item \textbf{查漏补缺：}从专用品、通用替代品、配套品和前置品四个角度检查是否遗漏了合理的商品类型，尤其关注配套品和前置品。只补充需求和图片能够支持的方案。
    \item \textbf{去除重复：}合并会检索到同一类商品的同义或近义关键词，但保留代表不同商品类型或解决方式的关键词。
    \item \textbf{修正偏差：}删除明显无法解决需求的关键词，并将过于宽泛的词改为能够指向具体商品的搜索词。不确定时优先保留。
\end{enumerate}

以补充和修正为主，不要为凑数强行扩展。所有关键词必须对应可购买的实物商品，表述应简洁、具体，并符合电商搜索习惯。最终列表最多包含六个关键词；超出时优先保留最能满足需求且能覆盖不同解决方式的关键词。

\medskip
\textbf{输入}\\
用户意图：\texttt{\{intent\}}\\
图片内容：\texttt{\{content\}}\\
用户提问：\texttt{\{query\}}\\
选品注意：\texttt{\{note\}}\\
上一轮关键词：\texttt{\{first\_round\_keywords\}}

\medskip
\textbf{输出}\\
仅输出最终关键词组成的合法 JSON 列表，不要输出复查过程或其他内容。

\end{promptbox}
\end{CJK*}

\subsection{Fine-Grained Annotation Prompt}

Below is an abridged version of the annotation prompt. We omit repeated
explanations and few-shot examples for space. The complete Chinese prompt will be released with the code.

\begin{CJK*}{UTF8}{gbsn}
\begin{promptbox}
\small

{\color{promptlabel}\bfseries System}\\
用户上传了一张图片并提出需求，希望购买商品来解决问题。给定一个候选商品，请判断其与用户需求的相关性等级。

\medskip
{\color{promptlabel}\bfseries User}\\
综合用户的图像内容、原始问题和推断意图，以及候选商品的标题、图描述和类目，判断商品是否能够满足用户需求。商品标题和图描述是主要依据，类目作为辅助；不要仅根据关键词重合判断相关性。

\smallskip
请按以下顺序判断：

\begin{enumerate}[nosep,leftmargin=*]
    \item 若商品与需求无实质关联，标为 \textbf{N3}；若商品与需求相关或表面相似，但不能解决或实质性帮助解决核心需求，标为 \textbf{N2}。
    \item 若商品原则上能够解决需求，但违反关键的安全、兼容性或适用性约束，标为 \textbf{N1}。
    \item 若商品能够直接、有效地满足需求并符合所有适用约束，标为 \textbf{L1}；若其能够解决或辅助解决需求，但在效果、适用性、专业性或完整性方面明显较弱，标为 \textbf{L2}。
\end{enumerate}

不要过度推测潜在风险。商品是否专用只是区分 L1 和 L2 的依据之一，而不是充分或必要条件。

\medskip
\textbf{输入}\\
图像内容：\texttt{\{content\}}\\
原始问题：\texttt{\{query\}}\\
用户意图：\texttt{\{intent\}}\\
商品标题：\texttt{\{title\}}\\
商品图描述：\texttt{\{product\_caption\}}\\
商品类目：\texttt{\{category\}}

\medskip
\textbf{输出}\\
仅输出合法 JSON：\\
\texttt{\{"grade":"L1|L2|N1|N2|N3","reason":"判断依据"\}}

\end{promptbox}
\end{CJK*}

\end{document}